# X-ray properties of head-tail radio sources in clusters of galaxies

A.C. Edge[1] & H. Röttgering[1,2,3]
[1]Institute of Astronomy, Madingley Road, Cambridge CB3 0HA, UK
[2]MRAO, Madingley Road, Cambridge CB3 OHE, UK
[3]Leiden Observatory, PO Box 9513, 2300 RA, Leiden, The Netherlands

**ABSTRACT**

From ROSAT imaging data we have detections and upper limits for a sample of 26 tailed radio sources in clusters of galaxies mostly from the sample of O'Dea & Owen (1985). All sixteen of the detected sources are unresolved in the ROSAT PSPC images. The sources bright enough to perform X-ray spectral analysis have power-law indices similar to BL Lacs and Seyfert galaxies. We find that there is a highly significant correlation between the core radio flux density and the X-ray flux but only a weak correlation between the total radio flux density and the X-ray flux. The trend is similar to that found in earlier studies of 3C radio galaxies with *Einstein* and more recently with ROSAT. The result adds an additional constraint on models for the unification of BL Lac objects with FR I radio sources. Also this result indicates that the observed enhanced X-ray emission near tailed sources is more likely to be due to nuclear emission rather than substructure in the extended cluster gas.

## 1 INTRODUCTION

Tailed radio sources have a tad-pole like asymmetric structure; its morphology is due to the movement through a dense inter cluster medium of the galaxy hosting the radio-jets (e.g. Miley 1980). This class of radio sources is commonly divided into narrow angle tailed sources (NATs) and wide angle sources (WATs); its main distinction being the size of the opening angle between the extremities of the two radio beams and the hosting galaxy. Studying the interaction between the radio jet and the cluster medium reveals not only information on the physics of the radio source, but also on the clusters they live in (e.g. O'Dea, Sarazin & Owen 1987). It has been argued that the main difference between BL Lac objects and FRI radio galaxies can be explained through orientation effects (Padovani & Urry 1990, 1991 and Urry, Padovani & Stickel 1991). Since head tailed radio galaxies are a relatively well understood sub-class of FRI radio galaxies, detailed studies of tailed radio galaxies will constrain models for the unification of BL Lac objects and FRI radio sources.

Extensive radio surveys of clusters have detected many head-tail radio galaxies (*e.g.* Owen *et al.* 1982, O'Dea & Owen 1985, Zhao, Burns & Owen 1989, Owen, White & Burns 1992 and Owen, White & Ge 1993). Recently Burns *et al.* (1994) found that cluster radio sources often have X-ray emission associated with them indicating that perhaps each radio galaxy is itself associated with a sub-cluster and providing evidence that the sharp bends that are observed in some of the jets of WATs (*e.g.* Eilek *et al.* 1984) are due to large scale motion in the intracluster medium near the radio source. Unfortunately the archival Einstein data used by Burns *et al.* (1994) was of relatively poor spatial resolution (1–1.5′) so the differentiation of unresolved and extended X-ray emission was difficult.

With the availability of archival ROSAT images of clusters that offer substantially better spatial resolution (30″) images, it is now possible to attempt a similar analysis to that of Burns *et al.* (1994). Using the sample of head-tail radio galaxies of O'Dea & Owen (1985) that were observed by ROSAT we wish to address the question of the origin of the excess X-ray emission near FR I radio sources in clusters.

## 2 RESULTS

We have taken 18 PSPC images and 1 HRI image from the ROSAT archive for 24 narrow-angle tail (NAT) sources (in 16 different clusters) of the O'Dea & Owen (1985) sample (hereafter O&O85). This represents 42 per cent of the original sample of 57. To this sample we added two further galaxies in A2634 (Eilek *et al.* 1984) and A3667 (Röttgering *et al.* 1995). The galaxy in A2634 is a classical Wide Angle Tail (WAT) so is not in the O&O85 sample and is included as a comparison to this class of cluster radio sources. The NAT galaxy in A3667 is at too low a declination to be observed by the VLA but is noted in Röttgering *et al.* (1995) as an X-ray source. One of the O&O85 sample (0309+411) is now known not to be in the originally proposed cluster, and has been identified with a giant radio source with a projected size of 1.8 Mpc and a quasar-like nucleus (de Bruyn 1989). We give some source parameters in tables 1 and 2 (see below), but we have not used this source in our subsequent analysis. Table 1 summarises the detections and upper limits corrected for galactic absorption and assuming a power-law index for the X-ray emission of 2.4 (the average spectral index found in the brighter sources, see below). Six sources have already been presented in the literature, IC 310 and NGC 1265 in Perseus (A426) by Rhee *et al.* (1994), 1615+351 and 1621+380 in poor Zwicky clusters by Feretti *et al.* (1994) and 3C40 in A194 and 3C264 in A1367 by Crawford & Fabian (1995). The limits and detections presented here are consistent with the published values. None of the detected sources is significantly extended beyond the spatial resolution of the PSPC (approximately $30''$ FWHM). Five of the detections were bright enough to perform X-ray spectral fitting. Table 2 gives the results for the best power law and thermal fits. Both are equally acceptable on a statistical basis. However, given the presence of compact radio nuclei and the lack of extended X-ray emission, the power-law spectrum is clearly physically preferred and we do not consider thermal models further. This illustrates the difficulty in differentiating between emission mechanisms even with moderately high signal-to-noise PSPC data. The spectrum of IC 310 is presented by Rhee *et al.* (1994) who suggests that IC310 may have a BL Lac nucleus from its X-ray spectral properties. The X-ray spectral analysis shows that the power-law spectral indices are similar to Seyfert galaxies (Turner, George & Mushotzky 1993, Brandt *et al.* 1994) and BL Lac objects (Fink *et al.* 1992) in the PSPC band.

Figures 1 & 2 show the total and core radio flux densities plotted against the PSPC or HRI flux. There is clearly a strong correlation between the core radio and the X-ray fluxes. Figure 2 illustrates that there is little relation between the diffuse, steep-spectrum radio emission and the compact X-ray emission.

The correlation in Figure 1 is strong and for any given X-ray flux there is a range of radio flux density of a factor of approximately five. The origin of this scatter could be related a number of factors. Firstly, intrinsic variability of the sources (Seilestad, Pearson & Readhead 1983) could be important as most radio observations were made 5–15 years ago whereas the X-ray observations were all made in the last 4 years and the majority of flat spectrum radio sources are believed to vary on decade timescales. Secondly, the choice of 1.4 GHz as the reference frequency is somewhat arbitrary and, as the spectral indices of the cores are mostly flat or inverted due to synchrotron self-absorption. It is possible that flux measurements in the optically thin part of the spectrum (10–20 GHz) may be more representative of the total radio power. Lastly, the use of ROSAT images to determine X-ray fluxes leaves the question of any excess X-ray absorption open, whether it occurs in the galactic core or in the central region of a cooling flow (White *et al.* 1991). Whatever the uncertainties, the observations give a correlation with a scatter of a factor of 3–4 implying that the overall radio-X-ray spectra of the cores of the galaxies have a very similar shape over a range of flux of almost two orders of magnitude.

The relation between core radio and X-ray fluxes is tight enough that it can be used to fairly accurately predict the X-ray flux for any other cluster radio galaxy. For instance, all FR I radio sources with a core brighter than 100 mJy at 1.4 GHz should be detected in the ROSAT survey (*i.e.* an X-ray flux greater than $5 \times 10^{-13}$ $\mathrm{erg\,cm^{-2}\,s^{-1}}$). Also this result also explains why the 'classic head-tail' radio galaxy, NGC1265, is a weak X-ray source. Its core radio flux density of 12 mJy is relatively faint compared to other head-tail sources and it is only marginally detected above the cluster emission in the PSPC images which either had a short exposure or were far off-axis and had a poor point-spread function (Rhee *et al.* 1994).

The correlation between X-ray and radio powers has been noted by a number of authors in various classes of radio sources as a luminosity-luminosity correlation. Worrall & Birkinshaw (1994) and Feretti *et al.* (1994) note a correlation between core radio power and non-thermal X-ray luminosity from ROSAT data for low redshift FR I radio galaxies. To link this into this paper, NGC 6251 (Birkinshaw & Worrall 1993) is plotted in Figure 1 and is consistent with the overall correlation. This implies that moderate power radio galaxies also follow the relation presented here. If this is the case then the conclusion of Birkinshaw & Worrall (1993) that the X-ray photon index of 2.8±0.6 they find is anomalously steep is not necessarily correct as the spectrum is consistent with the results for the sources in this paper (although the excess column in NGC 6251 of $1.3 \times 10^{21}$ $\mathrm{cm^{-2}}$ is larger than than any presented here). At higher radio powers, Fabianno *et al.* (1984) find the core radio luminosity of the brightest 3C galaxies is well correlated with the X-ray luminosity from *Einstein* observations. Recently Crawford & Fabian (1995) have extended this relation to include more distant 3CR galaxies using ROSAT observations. Crawford & Fabian (1995) conclude that broad-line radio galaxies (BLRGs) have more X-ray emission with respect to the radio that other radio galaxies. Also, Worrall *et al.* (1994) have presented ROSAT observations of two 3CR galaxies and conclude that core dominated 3CR quasars have more radio emission with respect to the X-ray. The advantage of plotting the correlation in flux is that the upper limits can be more clearly related to observational constraints.

## 3 DISCUSSION

There is an extensive debate in the literature on the potential links between FR I radio galaxies and BL Lac objects (see Padovani & Urry (1990,1991) and Urry, Padovani & Stickel (1991)). The host galaxies of both are predominantly ellipticals (Abraham, McHardy & Crawford 1991) and extended radio emission is found at similar levels in both (Antonucci & Ulvestad 1985). The space density of BL Lacs and FR I radio galaxies are consistent with the latter being the 'parent population' of BL Lac hosts. The observed properties of nuclear source depend on the angle at which the observer views the relativistic jet and the underlying energy spectrum being amplified.

In order to place the X-ray/radio properties of these cluster radio galaxies we have compared them with the radio and X-ray selected BL Lac objects discussed in Perlmann & Stocke (1993). Perlmann & Stocke (1993) illustrate the differences in the X-ray and radio properties of BL Lac depending on how they were selected. These differences are most prominent when the ratio of core to extended radio flux density is plotted against the ratio of X-ray to radio power (Figure 3). The X-ray selected BL Lacs have significantly higher X-ray flux compared to radio than radio selected objects. The X-ray selected objects also have a lower dispersion in the ratio of core to extended radio power. The cluster radio galaxies lie well below the radio selected BL Lacs in Figure 3 due to their much more prominent extended emission (which is the property they are selected on). However, the cluster radio sources have a very similar ratio of X-ray to core radio emission as the radio-selected BL Lacs and similar average extended radio powers (Perlmann & Stocke 1993). These two similarities by no means prove that the cluster radio sources are BL Lacs as their core radio powers are two to three orders of magnitude smaller than the radio selected BL Lacs (*i.e.* core flux densities of 10–100 mJy at redshifts of 0.02–0.05 compared to 1–3 Jy and $z =0.2$–0.5), they are highly suggestive.

Perlmann & Stocke (1993) speculate that there are a number of BL Lac objects that have been unrecognised in low frequency radio samples. The X-ray bright head-tail radio sources are potentially consistent with these 'missing' objects from their radio and X-ray properties. Evidence from optical studies holds a highly significant role in this question. Two of the most luminous sources in this sample, 0905-098 in A754 (Harris *et al.* 1984) and 3C264 in A1367 (Elvis *et al.* 1981) do not exhibit optical emission lines (the canonical optical property of a BL Lac), while two do, IC 310 in Perseus (Owen, Ledlow & Keel 1995) and 3C465 in A2634 (De Robertis & Yee 1990), the latter having a weak broad component to the H$_\alpha$ line (possible evidence against a BL Lac nucleus). One final line of evidence is the presence of an unresolved nucleus in optical imaging. This is a very difficult problem in the FR I radio galaxies as, scaling from radio-selected BL Lacs, the nucleus will be 3–4 magnitudes fainter than the host galaxy. This is entirely reversed for more distant BL Lac objects where the nucleus is brighter than the host. This problem has been discussed by Browne & Marchã (1993) in relation to X-ray selected BL Lacs where objects at redshifts less than 0.4 can become misclassified. Only one of the galaxies in this study has a published HST image, 3C264 (Crane *et al.* 1993). The HST image shows an unresolved nucleus and a $0.65''$ jet and subsequent NTT spectroscopy found no emission lines. The optical flux of the nucleus of 174 $\mu$Jy (Crane *et al.* 1993) is consistent with that expected from the radio/optical/X-ray spectrum of radio-selected BL Lacs given a core radio flux density of 250 mJy (Wolter *et al.* 1994). Clearly from just one case it is not possible to conclude that the nuclear source is a "mini-BL Lac" but HST imaging observations of many more FR I radio galaxies have either been made or are scheduled so situation will change soon. The correlation presented here, if applicable to the optical, would predict an optical nucleus with a flux correlated to the core radio flux density. Also the presence and length of any jet may provide a constraint on the viewing angle.

Burns *et al.* (1994) have recently published an analysis of Einstein images of clusters containing radio galaxies. They conclude that there is a very high fraction of X-ray 'substructure' in these images and that the radio galaxies lie very close to these 'substructures'. This association of radio galaxies with subclusters which may be merging with the parent cluster is an attractive one. However, the observation that many cluster radio sources are themselves X-ray sources (a possibility discussed in Burns *et al.*) draws us to a conclusion that much of the 'substructure' described by Burns *et al.* is due to unresolved X-ray sources.

The archival Einstein data used in the Burns *et al.* study has substantially poorer spatial resolution than the ROSAT data ($1.5'$ to $30''$) so the difference between point source and extended emission is more difficult to determine. Examining the radio/X-ray overlays in Burns *et al.* (1994) we find at least 8 sources that are apparently 'core-dominated' radio sources (*e.g.* A194, A415, A1367 and A2147). This number is probably an underestimate as the radio maps are printed at relatively limited resolution. When the additional observation that a number of the Burns *et al.* (1994) sources are associated with the brightest galaxy in the cluster (*i.e.* A133, Hydra-A and four others) which is almost invariably at the peak of the extended X-ray emission, then the relation between X-ray substructure and radio sources is undermined as *at least* 15 of the 25 sources within 150 kpc of an X-ray peak can be explained without the need for X-ray substructure. When these point sources and central galaxy sources are subtracted from the histograms in Figure 2 in Burns *et al.* (1994) then number of close associations of X-ray 'clumps' and radio galaxies is closer that expected from random ($<$10 observed, 5–7 expected). This is not to say that X-ray morphology and the presence of radio sources are unrelated. This is most dramatically illustrated in A2256 (Röttgering *et al.* 1994) and in A3667 (Röttgering *et al.* 1995) where strong X-ray substructure is found in clusters with very unusual diffuse radio emission. However, to prove a link between X-ray morphology and radio galaxies requires an X-ray flux-limited sample with homogeneous radio and X-ray imaging data.

Given the significance of the core radio emission, it is important to increase the number of core radio flux densities in the literature. Compared to the problems of measuring extended radio emission from aperture synthesis maps, core radio flux densities are relatively trivial to determine. In addition, more high frequency (10–20 GHz) radio images of these galaxies would help to establish the radio spectral index of the core and improve the contrast of the core to the extended emission.

## 4 CONCLUSIONS

The correlation between core radio and X-ray fluxes for cluster, head-tail radio sources provides an important insight into the geometry and physics of the core regions of these galaxies. Work on other radio galaxies, particularly ones selected in clusters, in the radio, X-ray and optical will allow these galaxies to be viewed in the framework of the possible unification of FR I radio sources and BL Lacs.

## ACKNOWLEDGMENTS

ACE acknowledges support from PPARC. HR acknowledges support from an EEC twinning project and funding from a programme subsidy granted by the Netherlands Organization for Scientific Research (NWO). This project would not have been possible without the Leicester ROSAT Archive and the efforts made by Steve Sembay, Julian Osbourne and Jeremy Ashley to keep it running so smoothly. We would also like to thank Carolin Crawford, Andy Fabian, Steve Allen and Dave White for useful discussions and suggestions. We thank the referee, Chris O'Dea, for his comments.

| Cluster | radio source name | IAU name | PSPC count rate ((count/s in Hard band, 0.4–2.4keV) | Unabsorbed 0.1–2.4 keV X-ray flux ($10^{-14}$ erg cm$^{-2}$ s$^{-1}$) | total radio flux density (Jy at 1.4GHz) | core radio flux density (Jy at 1.4 GHz) |
|---|---|---|---|---|---|---|
| A85 | | 0039-095A | $< 1.88 \times 10^{-3}$ | $< 7.3$ | 0.07 | 0.0035 |
| | | 0039-095B | $< 6.27 \times 10^{-3}$ | $< 24.2$ | 0.05 | 0.0156 |
| | | 0039-097 | $< 9.40 \times 10^{-3}$ | $< 36.4$ | 0.12 | 0.007 |
| A119 | | 0053-015 | $< 6.58 \times 10^{-3}$ | $< 26.0$ | 1.50 | 0.042 |
| | | 0053-016 | $5.26 \pm 1.64 \times 10^{-3}$ | $20.8 \pm 6.5$ | 1.13 | 0.008 |
| A194 | | 0123-016A | $6.76 \pm 1.02 \times 10^{-3}$ | $27.0 \pm 4.1$ | 0.91 | 0.017 |
| | 3C40 | 0123-016B | $17.6 \pm 1.22 \times 10^{-3}$ | $70.3 \pm 4.9$ | 4.27 | 0.060 |
| A401 | | 0255+133 | $< 3.22 \times 10^{-3}$ | $< 16.4$ | 0.70 | 0.009 |
| | | 0256+132 | $< 3.22 \times 10^{-3}$ | $< 16.4$ | 0.19 | 0.003 |
| A426 | | 0309+411 | $51.0 \pm 3.0 \times 10^{-3}$ | $266.1 \pm 15.7$ | 0.30 | 0.294 |
| | IC310 | 0313+411 | $85.0 \pm 2.0 \times 10^{-3}$ | $443.4 \pm 10.4$ | 0.67 | 0.130 |
| | NGC 1265 | 0314+416 | $10.0 \pm 4.0 \times 10^{-2}$ | $52.2 \pm 20.9$ | 7.67 | 0.012 |
| Zw | | 0335+096 | $2.46 \pm 1.23 \times 10^{-3}$ | $13.0 \pm 6.5$ | 0.37 | 0.017 |
| A496 | | 0431-134 | $4.24 \pm 1.11 \times 10^{-3}$ | $17.5 \pm 4.6$ | 0.52 | 0.024 |
| A754 | | 0905-098 | $128. \pm 7. \times 10^{-3}$ | $524.0 \pm 28.7$ | 0.37 | 0.119 |
| | | 0907-091 | $2.04 \pm 0.94 \times 10^{-3}$ | $8.4 \pm 3.9$ | 0.39 | 0.031 |
| A1367 | 3C264 | 1142+198 | $132. \pm 3. \times 10^{-3}$ | $453.9 \pm 10.3$ | 5.45 | 0.250 |
| A1656 | | 1256+282 | $< 9.02 \times 10^{-3}$ | $< 26.0$ | 0.13 | 0.002 |
| A1758 | | 1330+507 | $1.83 \pm 0.61 \times 10^{-3}$ | $5.9 \pm 2.0$ | 0.16 | 0.005 |
| A2142 | | 1556+274 | $< 1.06 \times 10^{-2}$ | $< 40.5$ | 0.09 | 0.021 |
| Zw | | 1615+351 | $7.50 \pm 1.08 \times 10^{-3}$ | $41.4 \pm 6.0$ | 0.20 | 0.032 |
| Zw | | 1621+380 | $< 12.0 \times 10^{-3\dagger}$ | $< 59.5$ | 0.18 | 0.021 |
| A2197 | | 1624+406 | $1.71 \pm 0.68 \times 10^{-3}$ | $5.0 \pm 2.0$ | 0.04 | 0.033 |
| A2319 | | 1918+439 | $< 4.64 \times 10^{-3}$ | $< 21.4$ | 0.08 | $<0.008$ |
| A2634 | 3C465 | | $13.2 \pm 2.0 \times 10^{-3}$ | $54.1 \pm 8.2$ | 7.87 | 0.150 |
| A3667 | | | $41.2 \pm 3.0 \times 10^{-3}$ | $178.1 \pm 13.0$ | 0.70 | 0.210 |

**Table 1.** PSPC and HRI count rates used in the analysis. The count rates marked with a dagger are from the HRI. All others are from PSPC data.

| Cluster | radio source | Power Law Index Temperature | Column density | Chi-squared |
|---|---|---|---|---|
| A426 | IC310 | $3.33^{+0.51}_{-0.43}$ | $2.6^{+0.8}_{-0.6} \times 10^{21}$ cm$^{-2}$ | 23.3 for 26 d.o.f. |
|  | 0313+411 | $0.77^{+0.17}_{-0.14}$ keV | $1.5^{+0.4}_{-0.3} \times 10^{21}$ cm$^{-2}$ | 25.7 for 25 d.o.f. |
| A426 | 0309+411 | $2.05^{+0.66}_{-0.53}$ | $1.9^{+1.2}_{-0.8} \times 10^{21}$ cm$^{-2}$ | 29.1 for 26 d.o.f. |
|  | Quasar | $2.56^{+4.54}_{-1.34}$ keV | $1.4^{+0.7}_{-0.5} \times 10^{21}$ cm$^{-2}$ | 29.6 for 25 d.o.f. |
| A754 | 0905–098 | $2.36^{+0.17}_{-0.16}$ | $7.9^{+1.0}_{-0.8} \times 10^{20}$ cm$^{-2}$ | 27.0 for 27 d.o.f. |
|  |  | $1.23^{+0.20}_{-0.21}$ keV | $5.9^{+0.7}_{-0.5} \times 10^{20}$ cm$^{-2}$ | 26.9 for 26 d.o.f. |
| A1367 | 3C264 | $2.45^{+0.13}_{-0.14}$ | $2.6^{+0.4}_{-0.4} \times 10^{20}$ cm$^{-2}$ | 31.1 for 26 d.o.f. |
|  | 1142+198 | $0.90^{+0.13}_{-0.14}$ keV | $1.2^{+0.2}_{-0.2} \times 10^{20}$ cm$^{-2}$ | 34.1 for 25 d.o.f. |
| A3667 |  | $2.48^{+0.45}_{-0.43}$ | $5.8^{+1.8}_{-1.6} \times 10^{20}$ cm$^{-2}$ | 20.0 for 25 d.o.f. |
|  |  | $1.10^{+0.60}_{-0.40}$ keV | $3.8^{+1.3}_{-0.9} \times 10^{20}$ cm$^{-2}$ | 21.2 for 24 d.o.f. |

**Table 2.** The spectral results for the brightest five sources from PSPC spectra. The errors are all at the 90% confidence level for one interesting parameter.

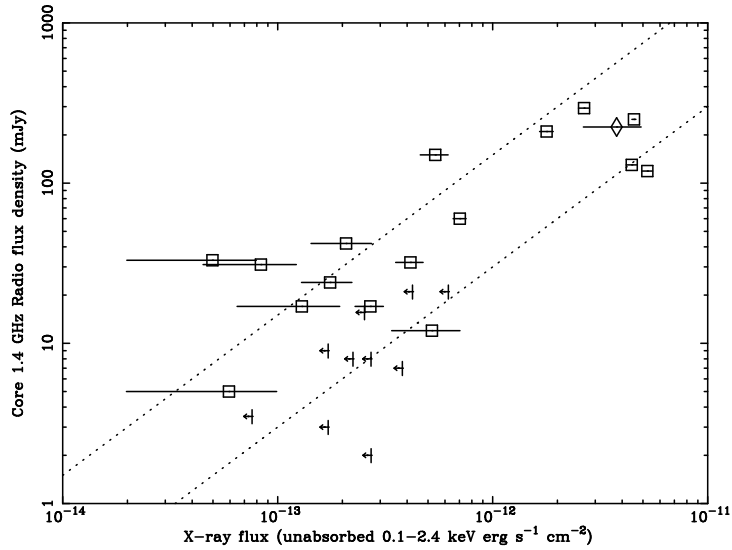

**Figure 1.** The core radio flux of the head-tail radio sources plotted against 0.1–2.4 keV X-ray flux (assuming a power law spectral index of 2.4 and galactic absorption). The diamond indicates the point for NGC 6251 from Birkinshaw & Worrall (1993).

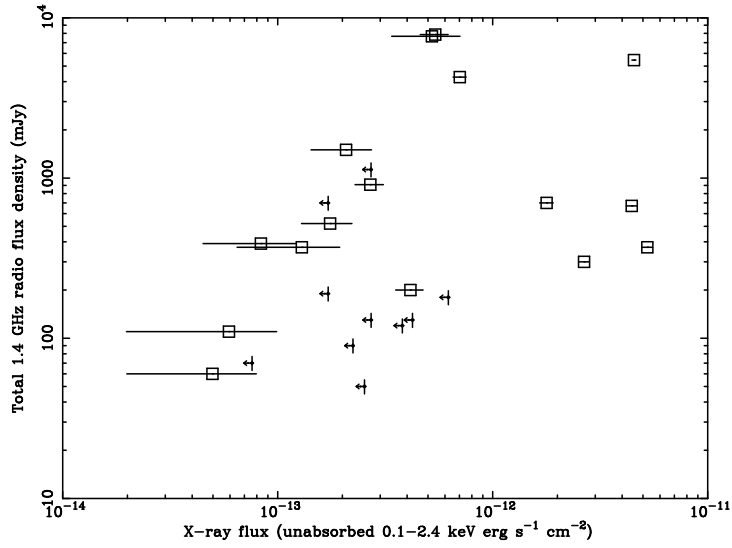

**Figure 2.** The total radio flux density of the head-tail source plotted against X-ray flux (with the same definitions as Figure 1).

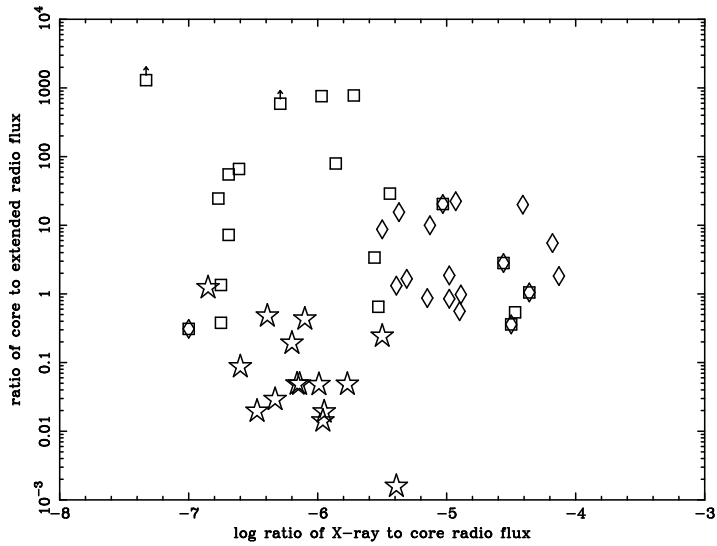

**Figure 3.** The ratio of total-to-core radio flux densities plotted against the ratio of X-ray to core radio flux density using the convention of Perlmann & Stocke (1993). The stars are the results from this paper (for detections only). The squares are radio-selected BL Lacs and the diamonds are X-ray-selected BL Lacs taken from Perlmann & Stocke. Several BL Lacs are both radio and X-ray-selected so appear twice. The cluster radio sources without an X-ray detection will lie mostly below 0.01.